\documentclass[twocolumn,showpacs,preprintnumbers,amsmath,amssymb]{revtex4-1}
\usepackage{graphicx}
\usepackage{dcolumn}
\usepackage{bm}
\usepackage{color}
\begin{document}
%
%
\title{Electric-field-induced stress relaxation in $\alpha$-phase poly(vinylidene fluoride) films}
\author{Xiangtong He$^{\dag}$, Shaopeng Pei$^{\dag}$, Robert E. Baier$^{\ddag}$, and John Y. Fu$^{\dag}$}
%
%
\affiliation{$^{\dag}$Department of Mechanical and Aerospace Engineering, The State University of New York, Buffalo, NY, 14260, USA\\$^{\ddag}$Department of Oral Diagnostic Sciences, The State University of New York, Buffalo, NY, 14214, USA}
%
%
%
%
\begin{abstract}
The relationship between elastic fatigue and electrical cyclic loading in $\alpha$-phase poly(vinylidene fluoride) films has been investigated. Our experimental studies have shown that the electric-field-induced fatigue behavior can be described by a stress relaxation, which belongs to the Kohlrausch function group, and the corresponding exponent is a modified two-parameter Weibull distribution function.
\end{abstract}
%
%
%
%
\maketitle
%
%
%
%

The non-Debye relaxation behavior has been observed in various condensed matter systems \cite{nondebye1987}. Many time-dependent materials properties are closely related to this behavior; one of them is the physical deterioration of materials. It has been known that elastic properties of certain materials would deteriorate after cyclic loading and can be well described by the following empirical stress relaxation function \cite{pierce1923}
\begin{equation}
E(t)=E_{s}\mathrm{exp}\left[-\left(\frac{t}{t_{p}}\right)^{1-k}\right],
\label{landau1937}
\end{equation}
where $E(t)$ is the time-dependent Young's modulus and $E_{s}$ the static Young's modulus; $t$ represents the time variable and $t_{p}$ the effective relaxation time; the exponent $k$ varies from 0.6 to 0.7 for different materials in practice. Clearly, this equation is the Kohlrausch function or the stretched exponential function \cite{kohlrausch}.

It might be interesting to examine whether the stress relaxation of dielectric materials, after electrical cyclic loading, obeys the Kohlrausch function or not. Due to the electromechanical coupling, either homogeneous coupling (piezoelectricity) or inhomogeneous coupling (flexoelectricity), dielectric materials usually undergo cyclic deformation when an ac electric field is applied to them. Thus, to some extent, electrical cyclic loading is equivalent to mechanical cyclic loading for dielectric materials. The coupling between electrical variables and mechanical variable is critical to maintaining the capability of storing electrical energy of a dielectric material. To prevent partial discharge channels or, more generally, dielectric-breakdown structure precursors from initiating and growing in a dielectric, the following mechanical force inequality must be satisfied \cite{zeller1984}: $\nabla_{r}W_{m}\geq\nabla_{r}W_{e}$; here $W_{m}$ represents the maximum mechanical energy beyond which dielectric-breakdown structure precursors will initiate and grow; $W_{e}$ is the electrostatic energy stored in the dielectric; $\nabla_{r}$ is the mathematical symbol representing the gradient with respect to the direction $r$. When the induced strain energy $W_{s}\ll W_{m}$ in the dielectric, $W_{s}$ must be equal to $W_{e}$ in equilibrium or two forces, which correspond to both $W_{s}$ and $W_{e}$, respectively, are balanced. Then, it can be easily seen that the electric strength of the dielectric is strongly dependent on its elastic properties. Obviously, if elastic properties of a dielectric deteriorate, its capability of storing electrical energy would also deteriorate. In our experimental studies, we are going to explore how the deterioration of elastic properties of a dielectric behaves under electrical cyclic loading. In the following text, the material preparation and the experiment procedures exploited in our studies will be discussed first.

\begin{figure}[h!]
\begin{center}
\includegraphics[width=0.7\columnwidth]{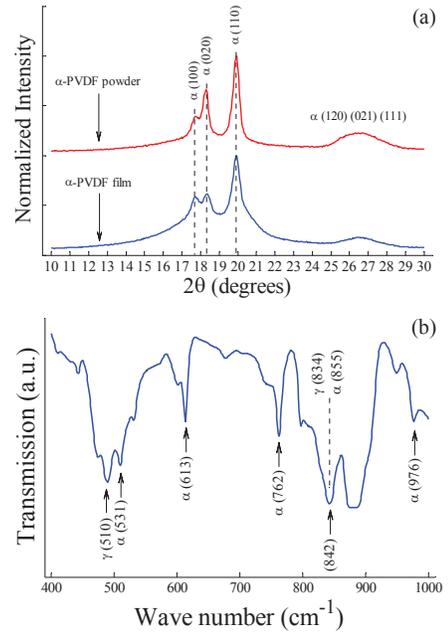}
\vspace{-0.10in}
\caption{(a) X-ray diffraction data of the commercial PVDF powder and the $\alpha$-phase PVDF film; (b) FTIR transmission spectrum of the $\alpha$-phase PVDF film.}
\vspace{-0.2in}
\end{center}
\end{figure}

Material and experimental procedures - in our studies, poly(vinylidene fluoride) [PVDF] polymer is chosen as the test material due to the following reasons: (a) PVDF and its copolymers usually have high dielectric breakdown strength values; (b) most of PVDF and its copolymers are low cost; (c) they are widely used in designing high energy density capacitors. Our PVDF powder was provided by Solvay Solexis and the procedure of preparing PVDF films is briefly summarized as follows: (1) Dissolving PVDF powder (SOLEF 1015/1001) in N-Methyl-2-pyrrolidone (NMP) solvent; the weight ratio of the powder to the solvent is kept as 1:10; (2) Stirring the PVDF-NMP solution on a magnetic stirring hot plate at a constant temperature of 60$^{\circ}$C degrees for approximately forty minutes; (3) Spin-coating a thin layer of the solution on a silicon wafer, previously cleaned with alcohol/DI water, by using a Laurell spin coater. The coating time and the coating rate (RPM) vary in different situations and are used to control the thickness of the layer; (4) Placing the spin-coated wafer in a heating oven at 180$^{\circ}$C degrees for $5\sim10$ minutes; (5) Removing the wafer from the oven and then immersing it in a mixture of ice and water for 20 minutes in order to interrupt the crystallization; (6) After such a quenching process, the wafer is placed back into the oven at 35$^{\circ}$C degrees for 5 hours for annealing; (7) Finally, leaving the wafer in DI water at room temperature for 24 hours and then the PVDF film can be peeled off from the surface of the wafer.

\begin{figure}[h!]
\begin{center}
\includegraphics[width=1.0\columnwidth]{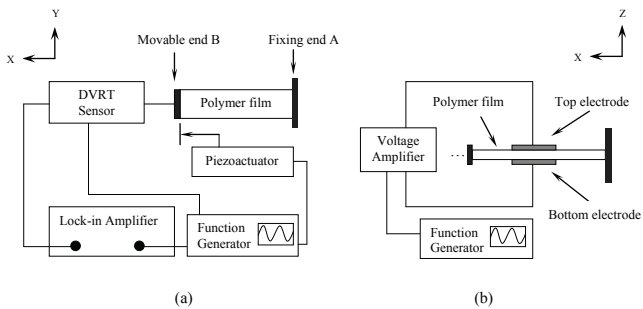}
\vspace{-0.10in}
\caption{(a) Lock-in detection setup for the Young's modulus measurement; (b) Experimental setup for electrical cyclic loading.}
\vspace{-0.2in}
\end{center}
\end{figure}

Our free standing PVDF films and the original PVDF powder were examined by X-ray diffraction and the results are shown in Fig. 1(a), from which we can see that the majority of crystalline phases in both the power and the film is $\alpha$ crystalline phase. The film was also examined by using a Perkin-Elmer 100 infrared spectrophotometer and the obtained Fourier transform infrared spectroscopy (FTIR) transmission spectrum is shown in Fig. 1(b). Based on the FTIR result, we can see that there is also a small quantity of $\gamma$ crystalline phase existing in the PVDF film. Using the method developed in Ref. \cite{gregorio1994}, we have estimated the fraction of crystalline phases of the considered PVDF film. Roughly speaking, the film is a semicrystalline polymer, in which the fraction of $\alpha$ phase is about $20\%\sim25\%$ and the rest are amorphous structures (here the fraction of $\gamma$ phase is neglected). In our experiment, the PVDF samples were neither stretched nor poled so that their crystalline phases (both $\alpha$ and $\gamma$ phases) are assumed to be randomly oriented in the spatial domain and, thus, we can approximately treat them as isotropic materials. Under this condition, the Young's modulus is regarded as the ratio of tensile stress over tensile strain.

\begin{figure}[h!]
\begin{center}
\includegraphics[width=1.0\columnwidth]{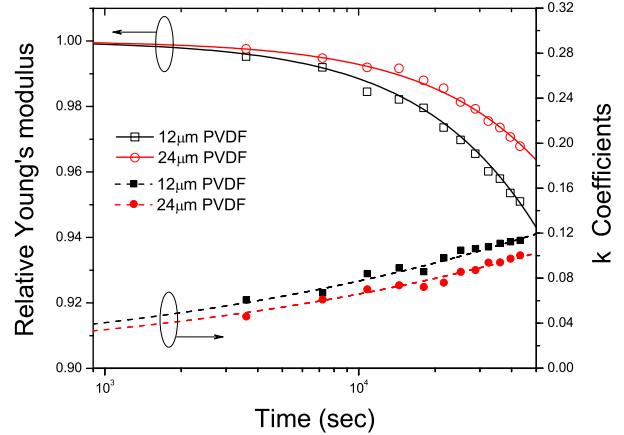}
\vspace{-0.10in}
\caption{Electric-field-induced elastic fatigue in PVDF films; open square and open circle represent the measured relative Young's moduli; solid square and solid circle represent the fitting coefficients $k$; solid curves are the corresponding fatigue fitting results that belong to the Kohlrausch function group; these two broken curves are the fitting results of $k$, which obey a modified two-parameter Weibull distribution function; the time variable given here represents the duration, throughout which electrical cyclic loading is applied.}
\vspace{-0.2in}
\end{center}
\end{figure}

The PVDF samples used in our Young's modulus measurement have rectangle shapes and are mounted on the moveable end B and the fixing end A, respectively, as shown in Fig. 2(a). To make sure that the corresponding Young's modulus values are measured within an elastic range, the lock-in detection of small deformation has been used. The induced strain inside a rectangle film can be generated by a tensile drawing tester, which is modified from a commercial piezoactuator. A low frequency sinusoidal signal provided by a function generator is used as both the control signal, which is sent to an amplifier to control the tester to create the periodic tensile deformation of the rectangle film, and the reference signal, which is sent to a lock-in amplifier, as shown in Fig. 2(a). The deformation at the B end is detected by using a MicroStrain DVRT sensor via the lock-in amplifier and is then used to calculate the induced tensile strain. The applied force exerted by the tester is calibrated by a load cell and is then used to calculate the tensile stress (here the cross-sectional area is defined as the product of the width and the thickness of the rectangle film).

The aforementioned electrical cyclic loading can be implemented by using the setup given by Fig. 2(b). An ac voltage is applied along the thickness direction that is perpendicular to the surface of the rectangle film as shown in Fig. 2(b). Here the voltage is designed to be a pure sinusoidal signal with zero dc offset to make sure that the film will not be poled during the corresponding electrical cyclic loading. In our experiment, the thickness of the rectangle film varies from $\mathrm{9\mu m}$ to $\mathrm{30\mu m}$; we therefore manipulate the applied voltage to make sure that the electric field inside different films is kept as a constant value of $\mathrm{0.1MV/cm}$. The measured Young's modulus value of the film that has not been exposed to external electric fields is regarded as $E_{s}$; whereas the modulus values of the films having undergone cyclic electrical loading are defined as $E(t)$ where $t$ is the duration, throughout which electrical cyclic loading is applied. The value of $E(t)$ of a rectangle film is designed to be immediately measured after the corresponding electrical cyclic loading is finished. If the measured $E(t)$ decreases with increasing the loading duration $t$, then $E(t)$ might be represented by a fitting relaxation function of $t$. We here use the relative Young's modulus, $E_{r}(t)=E(t)/E_{s}$, to define a normalized variation inasmuch as each rectangle film has a slightly different Young's modulus. Thus, our experimental investigation is to verify whether the relationship between $E_{r}$ and $t$ can be described by the Kohlrausch function or not. In other words, we would like to examine whether $E_{r}(t)=\mathrm{exp}[-(t/t_{p})^{1-k}]$ is held or not.

It is necessary to point out the difference between our proposed model and the Stark-Garton (SG) model \cite{stark1955}, which is a well known model to explain the electromechanical breakdown in polymers. The mathematical expression of SG model is: $\frac{1}{2}\varepsilon_{r}\varepsilon_{0}(\frac{V}{d})^{2}=ES_{e}$; here $\varepsilon_{0}$ and $\varepsilon_{r}$ represent the electric permittivity of free space and the relative permittivity of the test specimen - the polymer sample, respectively; $V$ is an external voltage and $d$ is defined as the resultant thickness of the sample after $V$ is applied; $E$ and $S_{e}$ represent the time-independent Young's modulus and the effective strain of the sample, respectively. Due to the plastic deformation induced in the vicinity of the electromechanical breakdown, SG model, in practice, is often modified as $\frac{1}{2}\varepsilon_{r}\varepsilon_{0}(\frac{V}{d})^{2}=K(S_{e})^{n}$ \cite{zhang2009}; here $K$ represents the strength index of the sample and $n$ varies from 0.1 to 0.6 for polymers \cite{zhang2009}. In either of these two models, $E$ or $K$ is regraded as the material parameter near the electromechanical breakdown; thus, both of them are assumed to be constant values. In our experimental studies, however, the applied electric field is kept as $\mathrm{0.1MV/cm}$ that is far away from the electromechanical breakdown limit. Thus, $E$ cannot be regarded as a time-independent value and how it is gradually altered by the duration $t$ is considered in this letter.

Results and discussion - the Young's moduli of our PVDF films have been measured and their average value is around $E_{s}=2.47\times10^{9} \mathrm{Nm^{-2}}$. Under the applied ac voltages of 10Hz, we measured $E_{r}$ of two polymer films with thickness values of $\mathrm{12\mu m}$ and $\mathrm{24\mu m}$, respectively. Our experimental studies have demonstrated that $E_{r}$ does indeed obey the Kohlrausch function of the electrical cyclic loading duration $t$ if $k$ is a modified two-parameter Weibull distribution, which is defined below.
\begin{equation}
k=\frac{1}{2}\left(1-\mathrm{exp}\left[-\left(\frac{t}{t_{wp}}\right)^{\beta}\right]\right),
\label{weibull}
\end{equation}
where $t_{wp}$ and $\beta$ are the scale parameter and the shape parameter, respectively, and $t$ here represents the electrical cyclic loading duration. For the film with thickness of $\mathrm{12\mu m}$, $t_{p}=1.25\times10^{6}$s, $t_{wp}=3.80\times10^{6}$s, and $\beta=0.3$; the minimum and maximum values of $k$ are 0.04 and 0.12, respectively. The fitting functions of $E_{r}$ (the elastic fatigue of $E_{r}$) and $k$ can be seen in Fig. 3. For the film with thickness of $\mathrm{24\mu m}$, the corresponding parameters are $t_{p}=1.96\times10^{6}$s, $t_{wp}=6.80\times10^{6}$s, and $\beta=0.3$; the minimum and maximum values of $k$ are 0.03 and 0.10, respectively. The corresponding fitting functions have also been shown in Fig. 3. It can be easily seen that the elastic fatigue of the $\mathrm{12\mu m}$ film is more severe, which corresponds to larger $k$ values. This suggests that $k$ might be related to the fraction of disordered structures of the film and would increase as the duration $t$ increases. To further explore the physical meaning of $k$, we placed the film with the deteriorated elasticity into a heating oven at 70$^{\circ}$C for 6 hours. After the annealing, we re-measured its Young's modulus and found that its $E_{s}$ can be recovered. We believe that the disordered structures represented by $k$ should be a reversible microstructural transformation caused by electrical cyclic loading. Finally, we must emphasize that $E_{r}$ is strongly frequency dependent in polymer materials. With the applied voltages of different frequencies, we would obtain different $E_{r}$ values. However, we found that the elastic fatigue of $E_{r}(t)$ can still be described by the Kohlrausch function via appropriate parameters ($t_{p}$, $t_{wp}$, and $\beta$).

Concluding remarks - the experimental evidence that the electric-field-induced elastic fatigue of PVDF films obeys the Kohlrausch function is reported here. It has also been observed that the corresponding exponent is a modified Weibull distribution function. Our studies may provide a new route to exploring the electric-field-induced fatigue failure in polymers.

\begin{flushleft}
\textbf{\small Acknowledgment}
\end{flushleft}
The authors sincerely thank Solvay Solexis for providing the PVDF powder used in their studies.
%
%
%
%
%

%
%
%
\end{document}